\documentclass[english,pop,showpacs,reprint,superscriptaddress]{revtex4-1}
\usepackage[T1]{fontenc}
\usepackage[latin9]{luainputenc}
\usepackage{float}
\usepackage{amsmath}
\usepackage{amssymb}
\usepackage{graphicx}
\usepackage{color}
\usepackage{soul}
\makeatletter

\usepackage{hyperref}

\makeatother

\usepackage{babel}
\begin{document}

\title{On the structure of the drifton phase space and its relation to the
Rayleigh\textendash Kuo criterion of the zonal-flow stability}

\author{Hongxuan Zhu}

\affiliation{Department of Astrophysical Sciences, Princeton University, Princeton,
NJ, 08544 }

\affiliation{Princeton Plasma Physics Laboratory, Princeton, NJ 08543}

\author{Yao Zhou}

\affiliation{Princeton Plasma Physics Laboratory, Princeton, NJ 08543}

\author{I. Y. Dodin}

\affiliation{Department of Astrophysical Sciences, Princeton University, Princeton,
NJ, 08544 }

\affiliation{Princeton Plasma Physics Laboratory, Princeton, NJ 08543}
\begin{abstract}
The phase space of driftons (drift-wave quanta) is studied within
the generalized Hasegawa--Mima collisionless-plasma model in the
presence of zonal flows. This phase space is made intricate by the
corrections to the drifton ray equations that were recently proposed
by Parker {[}J. Plasma Phys. \textbf{82}, 595820602 (2016){]} and
Ruiz \textit{et al.} {[}Phys. Plasmas \textbf{23}, 122304 (2016){]}.
Contrary to the traditional geometrical-optics (GO) model of the drifton
dynamics, it is found that driftons can be not only trapped or passing,
but they can also accumulate spatially while experiencing indefinite
growth of their momenta. In particular, it is found that the Rayleigh\textendash Kuo
threshold known from geophysics corresponds to the regime when such
``runaway'' trajectories are the only ones possible. On one hand,
this analysis helps visualize the development of the zonostrophic
instability, particularly its nonlinear stage, which is studied here
both analytically and through wave-kinetic simulations. On the other
hand, the GO theory predicts that zonal flows above the Rayleigh\textendash Kuo
threshold can only grow; hence, the deterioration of intense zonal
flows cannot be captured within a GO model. In particular, this means
that the so-called tertiary instability of intense zonal flows cannot
be adequately described within the quasilinear wave kinetic equation,
contrary to some previous studies. 
\end{abstract}
\maketitle

\section{Introduction}

The interaction between zonal flows (ZFs) and drift-wave (DW) turbulence
has a substantial effect on turbulent transport in fusion devices,
and hence has been actively studied in plasma physics for decades
\cite{Diamond2005review,Fujisawa09,Lin98,Biglari90,Dorland00,Jenko00,Connaughton2015}.
One of the popular reduced models of DW dynamics in ZFs is the wave
kinetic equation (WKE) \cite{Trines2005,Malkov2001,Kim2002,Vedenov1967,Kaw2001,Singh2014,Sasaki2018,Smolyakov1999,Smolyakov2000,Smolyakov2000prl},
which assumes the geometrical-optics (GO) approximation, i.e., loosely
speaking, that the DW wavelengths are vanishingly small compared to
ZF scales. Within this approximation, DWs can be understood as a gas
of ``driftons'', which are quasi-particles described by coordinates
$\boldsymbol{x}$, momenta $\boldsymbol{p}$ (DW wave vectors), and
energies $\mathcal{H}$ (DW frequencies). In particular, $\mathcal{H}=\mathcal{H}(t,\boldsymbol{x},\boldsymbol{p})$
serves as the quasi-particle Hamiltonian that determines the drifton
trajectory for given $\boldsymbol{x}$ and $\boldsymbol{p}$ at time $t$.
This model facilitates understanding of many important effects, including
the zonostrophic instability (ZI), i.e., the formation of ZF out of
DW turbulence \cite{Smolyakov2000,Smolyakov2000prl,Srinivasan2012,Parker2013,Parker2014}. 

It was shown recently that the drifton Hamiltonian used in previous
studies is oversimplified, and an improved Hamiltonian has been proposed
in Refs.~\cite{Parker2016,Ruiz2016} based on the generalized Hasegawa\textendash Mima
equation (gHME) \cite{krommes2000}. The corresponding improved WKE
(iWKE) accounts for the loss of drifton enstrophy to ZFs, and hence
is a more adequate GO model. The advantages of the iWKE are demonstrated
in Refs.~\cite{Parker2016,Ruiz2016} by numerical simulations. A
numerical comparison between the iWKE and the quasilinear gHME was
reported in Ref.~\cite{Parker2018}. However, it is also insightful
to explore the single-particle drifton dynamics, i.e., the drifton
phase-space trajectories in a prescribed ZF. Such study can help elucidate
the importance of individual terms in the drifton Hamiltonian. It
can also help us understand the nonlinear dynamics of ZFs, including
the nonlinear stage of the ZI, and identify factors that are important
for its saturation. Here, we report such study, which explores in
depth the drifton dynamics within the iWKE proposed in Refs.~\cite{Parker2016,Ruiz2016}.
Some of our results were also highlighted in Ref.~\cite{Zhu2017}.
The purpose of the present paper is to expand the discussion and to
elaborate on details.

Our main findings are as follows. (i) Contrary to the traditional
WKE (tWKE) of the drifton dynamics, which predicts \cite{Kaw2001,Singh2014,Sasaki2018}
nonlinear structures $\textit{à la}$ Bernstein\textendash Greene\textendash Kruskal
(BGK) waves \cite{BGK1957}, the iWKE predicts that driftons do not
have to be just passing or trapped. Instead, they can accumulate in
certain spatial locations while experiencing indefinite growth of
their momenta. We call such trajectories ``runaway''. (ii) Depending
on the ZF parameters, the drifton phase space can have three different
regimes. In Regime~1, passing, trapped, and runaway trajectories
coexist. In Regime~2, passing trajectories disappear entirely, but
both trapped and runaway trajectories can coexist. In Regime~3, only
runaway trajectories are left. (iii) Remarkably, Regime~3 is 
precisely the regime when the ZF amplitude exceeds the Rayleigh--Kuo
threshold known from geophysics \cite{Kuo1949}. Also notably, this
regime is not captured by the tWKE. (iv) We apply our phase-space
analysis to visualize the development of the ZI, particularly its
nonlinear stage, using both theoretical arguments and iWKE simulations.
Moreover, we find that the GO theory predicts that ZFs above the Rayleigh--Kuo
threshold can only grow; hence, the deterioration of intense ZFs cannot
be captured within a GO model \cite{Zhu2017,Zhu2018}. In particular,
this means that the so-called tertiary instability of intense ZFs
cannot be adequately described within a quasilinear WKE (including
the tWKE and the iWKE, which both assume the GO limit), contrary to some previous studies. Our
results serve as a stepping stone toward revising basic physics of
DW\textendash ZF interactions from the new perspective of drifton
phase-space dynamics beyond the traditional (tWKE-based) approach.

The rest of the paper is organized as follows. In Sec.~\ref{sec:section2},
the gHME and the iWKE are introduced. In Sec.~\ref{sec:section3},
the three different regimes of drifton phase-space structure are described.
The two critical ZF magnitudes that separate these three regimes are
also given. In Sec.~\ref{sec:section4}, the nonlinear ZI and the
TI are discussed. Our main conclusions are summarized in Sec.~\ref{sec:5}.
Auxiliary calculations are given in Appendix \ref{appendixA}.

\section{Basic Equations}

\label{sec:section2}

\subsection{The generalized Hasegawa\textendash Mima model}

\begin{figure}
\includegraphics[width=0.7\columnwidth]{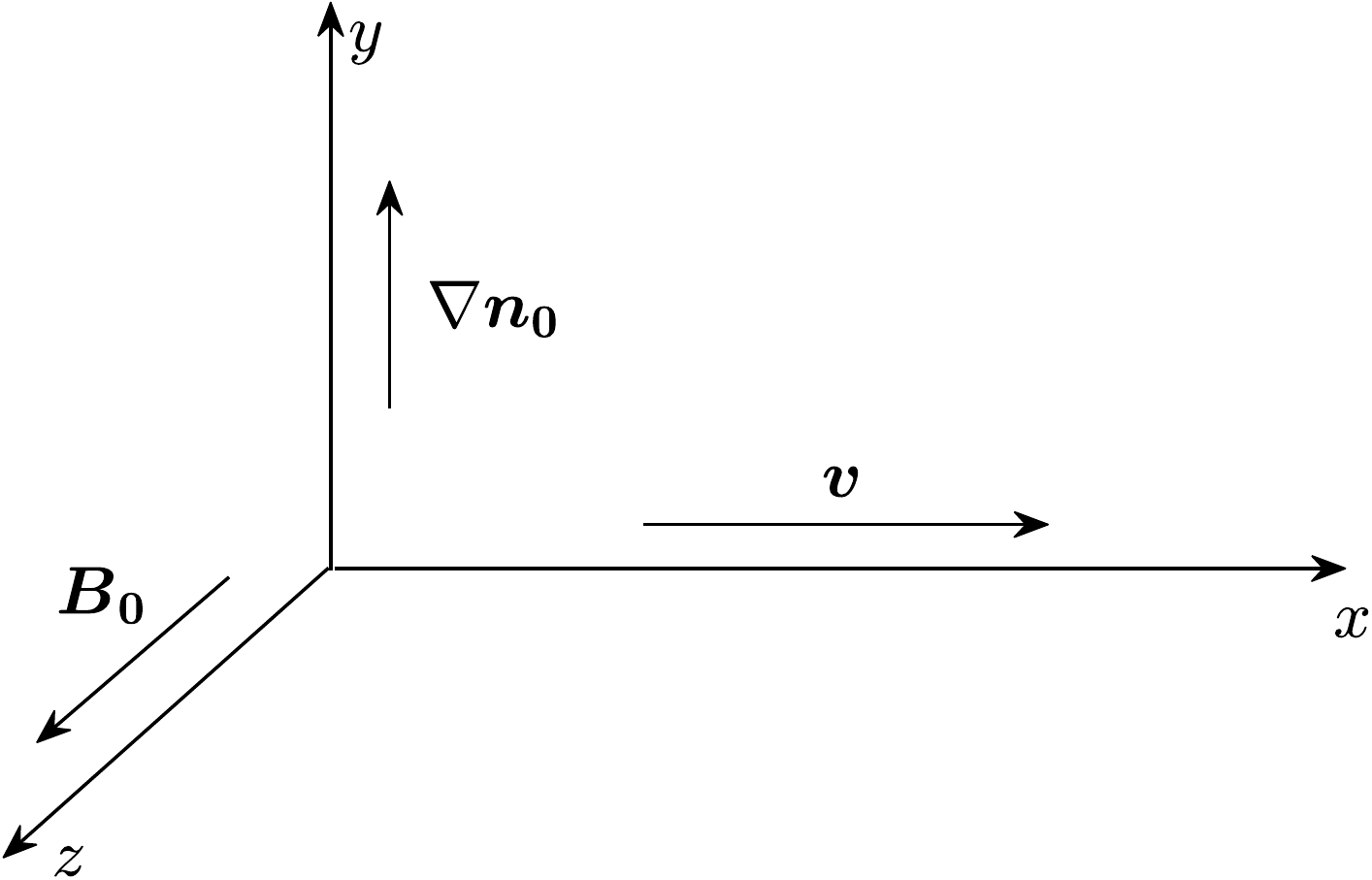}

\caption{The assumed coordinate system. Here, $\boldsymbol{B}_{0}$ is the
magnetic field, $n_{0}$ is the background electron density, and $\boldsymbol{v}$
is the ZF velocity.\label{fig:II_configuration}}
\end{figure}

First, let us introduce the original Hasegawa\textendash Mima equation
\cite{Hasegawa1977}. Consider a collisionless plasma in a uniform
magnetic field $\boldsymbol{B}_{0}$ in the $z$ direction, with the
equilibrium gradient of the background electron density $n_{0}$ in
the $y$ direction (Fig.~\ref{fig:II_configuration}). Ions are assumed
cold, while electrons are assumed to have a finite temperature $T_{e}$.
Suppose that perturbations to the electric field $\boldsymbol{E}$
are electrostatic, $\boldsymbol{E}=-\nabla\delta\varphi$, where $\delta\varphi(t,\boldsymbol{x})$
is the corresponding electrostatic potential on the two-dimensional
plane $\boldsymbol{x}\doteq(x,y)$. The electron response to $\boldsymbol{E}$
is adiabatic (yet see below), while the ion response can be described
by the $\boldsymbol{E}\times\boldsymbol{B}_{0}$ drift and the polarization
drift. Then, assuming the quasi-neutrality condition, the evolution
of $\delta\varphi$ is described by
\begin{multline}
\frac{\partial}{\partial t}\left[(\rho_{s}^{2}\nabla^{2}-1)\delta\varphi\right]\\
+\boldsymbol{u}_{E}\cdot\nabla\left[(\rho_{s}^{2}\nabla^{2}-1)\delta\varphi\right]+V_{*}\,\frac{\partial\delta\varphi}{\partial x}=0.\label{eq:II_hme_pre}
\end{multline}
Here, $\rho_{s}\doteq c_{s}/\Omega_{i}$ is the ion sound radius (we
use $\doteq$ to denote definitions), $c_{s}\doteq\sqrt{ZT_{e}/m_{i}}$
is the ion sound speed, $Z$ is the ion charge number, $\Omega_{i}\doteq Z|e|B_{0}/m_{i}$
is the ion gyrofrequency, $e$ is the electron charge, $\boldsymbol{u}_{E}\doteq\hat{\boldsymbol{z}}\times\nabla\delta\varphi/B_{0}$
is the $\boldsymbol{E}\times\boldsymbol{B}_{0}$ velocity, $\hat{\boldsymbol{z}}$
is the unit vector along the $z$ axis, $V_{*}\doteq T_{e}/(L_{n}B_{0}|e|)$
is the electron diamagnetic drift velocity, and $L_{n}\doteq(-\partial\ln n_{0}/\partial y)^{-1}$
is the characteristic length scale of $n_{0}$. Also, $\nabla^{2}\doteq\partial^{2}/\partial x^{2}+\partial^{2}/\partial y^{2}$
is the Laplacian.

Let us measure time in units $1/\Omega_{i}$ and length in units $\rho_{s}$.
Let us also introduce a normalized potential $\varphi\doteq e\delta\varphi/T_{e}$
and a normalized ``generalized vorticity'' $w\doteq(\nabla^{2}-1)\varphi$.
Then, Eq.~(\ref{eq:II_hme_pre}) can be written in the following
dimensionless form:
\begin{equation}
\frac{\partial w}{\partial t}+(\hat{\boldsymbol{z}}\times\nabla\varphi)\cdot\nabla w+\beta\,\frac{\partial\varphi}{\partial x}=0,\label{eq:II_hme}
\end{equation}
where $\beta\doteq V_{*}/c_{s}$ is treated as a (positive) constant.
Equation (\ref{eq:II_hme}) represents the original Hasegawa\textendash Mima
equation. 

Let us introduce the zonal average as $\langle f\rangle\doteq\int_{0}^{L_{x}}fdx/L_{x}$,
where $L_{x}$ is the system length in the $x$ direction. Then, perturbations
governed by Eq.~(\ref{eq:II_hme_pre}) include ZFs and DWs. The former
are identified as zonal-averaged perturbations, and the latter are
identified as fluctuations with zero zonal average. Strictly speaking,
electrons respond differently to ZFs and DWs. To account for this
and thus make the plasma model more realistic, the governing equations
can be rewritten as follows:
\begin{gather}
\frac{\partial w}{\partial t}+(\hat{\boldsymbol{z}}\times\nabla\varphi)\cdot\nabla w+\beta\,\frac{\partial\varphi}{\partial x}=0,\label{eq:II_ghme}\\
w=(\nabla^{2}-\hat{a})\varphi,\label{eq:II_ghme2}
\end{gather}
where $\hat{a}$ is an operator such that $\hat{a}=1$ for DWs and
$\hat{a}=0$ for ZFs \cite{Dorland1993Thesis,Hammett1993}. Equations
(\ref{eq:II_ghme}) and (\ref{eq:II_ghme2}) constitute the so-called
gHME \cite{krommes2000}, which is the model that we assume below.

\subsection{The improved WKE}

The ZF can be described using the average velocity $U(y,t)\doteq-\partial_{y}\langle\varphi\rangle$,
and DWs can be described using the zonal average of their Wigner function,
\begin{equation}
W(y,\boldsymbol{p},t)\doteq\langle\int d^{2}s\,e^{-i\boldsymbol{p}\cdot\boldsymbol{s}}\tilde{w}(\boldsymbol{x}+\frac{\boldsymbol{s}}{2},t)\tilde{w}(\boldsymbol{x}-\frac{\boldsymbol{s}}{2},t)\rangle,
\end{equation}
which in the GO limit can be understood as the drifton phase-space
distribution \cite{Ruiz2016}. (The discussion on the positive definiteness
of the Wigner function in the concept of quantum mechanics can be
found in Ref.~\cite{cartwright1976}.) The GO limit itself is defined
as the regime where 
\begin{equation}
\epsilon\doteq\max\left(\frac{\lambda_{\text{DW}}}{\lambda_{\text{ZF}}},\frac{\rho_{s}}{\lambda_{\text{ZF}}}\right)\ll1,
\end{equation}
where $\lambda_{\text{ZF}}$ and $\lambda_{\text{DW}}$ are the wavelengths
of ZFs and DWs, respectively. To proceed, the quasilinear approximation is used \cite{Srinivasan2012,Parker2013,Herring1963}, which assumes that the DW
self-interactions can be ignored. (Recent work \cite{Ruiz2018}
has also gone beyond the quasilinear approximation.) Then, the evolution
equations for $W$ and $U$ are \cite{Parker2016,Ruiz2016}
\begin{gather}
\frac{\partial W}{\partial t}=\{\mathcal{H},W\}+2\Gamma W,\label{eq:II_WKE_DW}\\
\frac{\partial U}{\partial t}=\frac{\partial}{\partial y}\int\frac{d^{2}p}{(2\pi)^{2}}\,\frac{p_{x}p_{y}W}{p_{D}^{4}}\,,\label{eq:II_WKE_ZF}
\end{gather}
where $p_{D}^{2}\doteq1+p_{x}^{2}+p_{y}^{2}$, and $\{\cdot,\cdot\}$
is the canonical Poisson bracket, namely,
\begin{equation}
\{A,B\}\doteq\frac{\partial A}{\partial\boldsymbol{x}}\cdot\frac{\partial B}{\partial\boldsymbol{p}}-\frac{\partial A}{\partial\boldsymbol{p}}\cdot\frac{\partial B}{\partial\boldsymbol{x}}\,.
\end{equation}
The Hermitian and anti-Hermitian parts of the Hamiltonian are given
by
\begin{gather}
\mathcal{H}=-\beta p_{x}/p_{D}^{2}+p_{x}U+p_{x}U''/p_{D}^{2},\label{eq:II_WKE_hamiltonian}\\
\Gamma=-U'''p_{x}p_{y}/p_{D}^{4},\label{eq:II_WKE_dissipation}
\end{gather}
where primes denote derivatives with respect to $y$. 

Equation (\ref{eq:II_WKE_DW}) is the iWKE as described in Refs.~\cite{Parker2016,Ruiz2016}.
In comparison, the tWKE used in previous studies is given by the same
Eq.~(\ref{eq:II_WKE_DW}) but with different $\mathcal{H}$ and $\Gamma$,
namely, 
\begin{gather}
\mathcal{H}_{t}=-\beta p_{x}/p_{D}^{2}+p_{x}U,\label{eq:II_tWKE_hamiltonian}\\
\Gamma_{t}=0,\label{eq:II_tWKE_dissipation}
\end{gather}
where the subscript ``$t$'' stands for ``traditional''. The iWKE
conserves the total enstrophy (per unit length in $x$) $Z_{\text{total}}=Z_{\text{DW}}+Z{}_{\text{ZF}}$
and the total energy (per unit length in $x$) $E_{\text{total}}=E_{\text{DW}}+E_{\text{ZF}}$
of the ZF\textendash DW system, where 
\begin{gather}
Z_{\text{DW}}\doteq\frac{1}{2}\int\frac{d^{2}p\,dy}{(2\pi)^{2}}\,W,\label{eq:II_enstrophy_dw}\\
Z_{\text{ZF}}\doteq\frac{1}{2}\int dy(U')^{2},\label{eq:II_enstrophy_zf}\\
E_{\text{DW}}\doteq\frac{1}{2}\int\frac{d^{2}p\,dy}{(2\pi)^{2}}\,\frac{W}{p_{D}^{2}},\label{eq:II_energy_dw}\\
E_{\text{ZF}}\doteq\frac{1}{2}\int dy\,U^{2}.\label{eq:II_energy_zf}
\end{gather}
In contrast, the tWKE conserves only the DW enstrophy but not the
total enstrophy, and as a result, can be unsatisfactory in many respects \cite{Parker2016,Ruiz2016}.

\subsection{Single-particle drifton dynamics}

\label{subsec:section2.3}

As an integral of the Wigner function over the whole phase space,
the DW enstrophy {[}Eq.~(\ref{eq:II_enstrophy_zf}){]} can be considered
as the total number of driftons \cite{Ruiz2016}. Due to the presence
of nonzero $\Gamma$ {[}Eq.~(\ref{eq:II_WKE_dissipation}){]}, $Z_{\text{DW}}$
is not conserved, hence the iWKE {[}Eq.~(\ref{eq:II_WKE_DW}){]}
does not conserve the total number of driftons. However, it can be
made conservative in the case of stationary $U$ by introducing $F(y,\boldsymbol{p},t)\doteq W(y,\boldsymbol{p},t)/[\beta-U''(y)]$;
then, Eq.~(\ref{eq:II_WKE_DW}) becomes \cite{Parker2016,Wordsworth2009}
\begin{equation}
\frac{\partial F}{\partial t}=\{\mathcal{H},F\},\label{eq:II_WKE_Parker}
\end{equation}
where $\mathcal{H}$ is still given by Eq.~(\ref{eq:II_WKE_hamiltonian}).
This shows that $F$ is conserved along drifton trajectories, which
are given by Hamilton's equations:
\begin{eqnarray}
\frac{dy}{dt} & = & \frac{\partial\mathcal{H}}{\partial p_{y}}=\frac{2p_{x}p_{y}}{p_{D}^{4}}(\beta-U''),\label{eq:II_ray_velocity}\\
\frac{dp_{y}}{dt} & = & -\frac{\partial\mathcal{H}}{\partial y}=-\frac{p_{x}}{p_{D}^{2}}(U'''+p_{D}^{2}U').\label{eq:II_ray_acceleration}
\end{eqnarray}
Equations (\ref{eq:II_ray_velocity}) and (\ref{eq:II_ray_acceleration})
describe the drifton dynamics within the iWKE. Since $\mathcal{H}$
does not depend on $x$, $p_{x}$ is also conserved along the trajectory.
Therefore, the drifton dynamics can be studied on the ($y,p_{y}$)
plane with $p_{x}$ serving as a parameter.

In more general situations where $U$ is not stationary, Eq.~(\ref{eq:II_WKE_Parker})
does not apply. But even in those situations, one can still view Eqs.~(\ref{eq:II_ray_velocity})
and (\ref{eq:II_ray_acceleration}) as equations of the drifton motion,
while $\Gamma$ only affects the evolution of the drifton density
along such trajectories, not the trajectories themselves.

\section{Drifton phase-space trajectories}

\label{sec:section3}

\begin{figure}
\includegraphics[width=1\columnwidth]{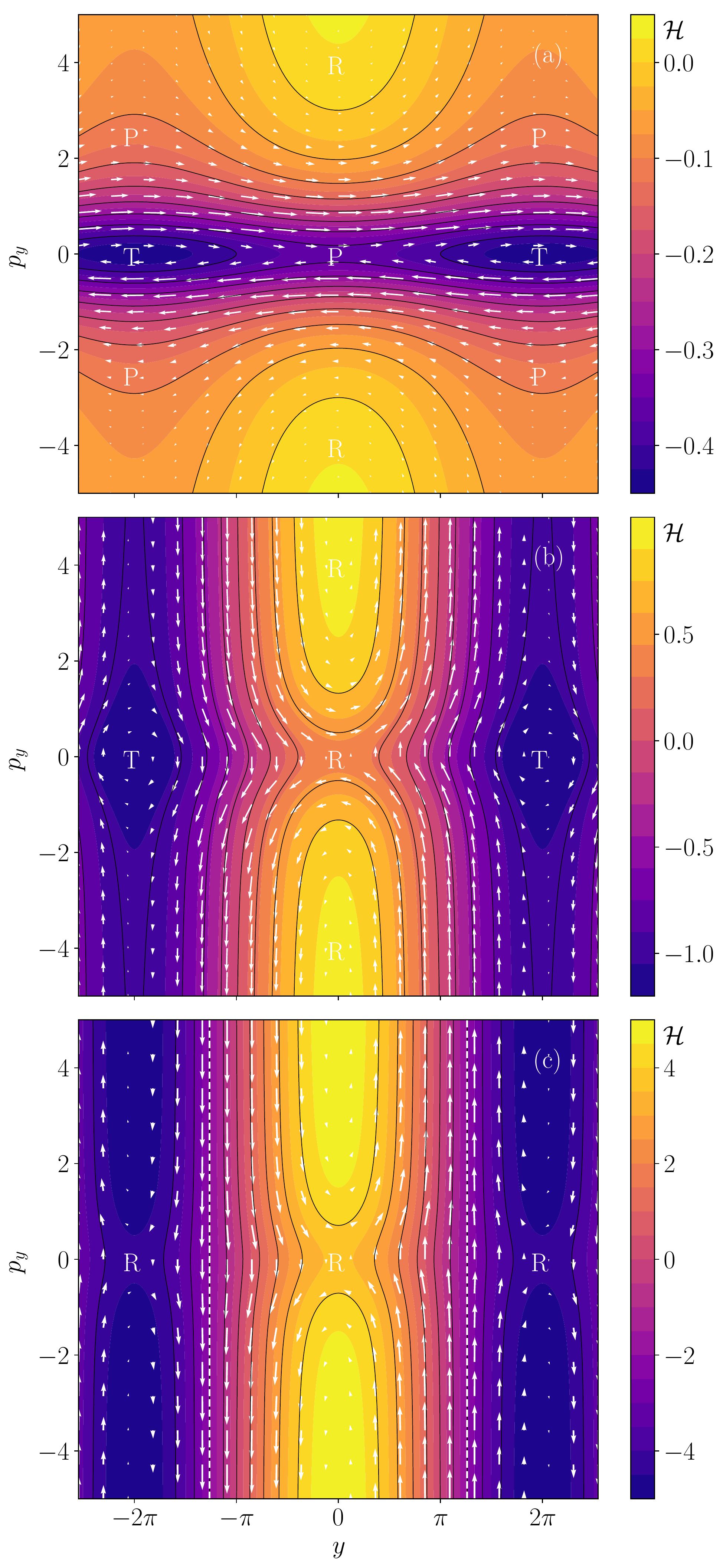}

\caption{Contour plots of $\mathcal{H}$ given by Eq.~(\ref{eq:III_H}). The
arrows show the phase-space velocity fields $(\dot{y},\dot{p_{y}})$
given by Eqs.~(\ref{eq:III_velocity}) and (\ref{eq:III_acceleration}).
Three different regimes are shown: (a) Regime~1, which corresponds
to a weak ZF ($u_{0}=0.1$); (b) Regime~2, which corresponds to a
moderate ZF ($u_{0}=2$); and (c) Regime~3, which corresponds to
a strong ZF ($u_{0}=10$). The labels ``T'', ``P'', and ``R''
denote trapped, passing, and runaway trajectories, respectively. The
white dashed lines in (c) are $|y|=y_{*}$, where $y_{*}$ is given
by Eq.~(\ref{eq:III_ystar}). In all cases, the parameters are $\beta=1$,
$q=0.5$, and $p_{x}=0.5$.\label{fig:III_traj}}
\end{figure}

For simplicity, we assume a sinusoidal ZF, namely, 
\begin{equation}
U(y)=u_{0}\cos qy,\label{eq:III_U_profile}
\end{equation}
where $u_{0}$ and $q$ are constant (for clarity, we assume $u_{0}>0$
and $q>0$). Then, Eq.~(\ref{eq:II_WKE_hamiltonian}) leads to the
following expression for the Hamiltonian: 
\begin{equation}
\mathcal{H}(y,p_{y})=-\frac{\beta p_{x}}{p_{D}^{2}}+p_{x}u_{0}\cos qy\left(1-\frac{q^{2}}{p_{D}^{2}}\right),\label{eq:III_H}
\end{equation}
and Eqs.~(\ref{eq:II_ray_velocity}) and (\ref{eq:II_ray_acceleration})
become
\begin{gather}
\frac{dy}{dt}=\frac{2p_{x}p_{y}}{p_{D}^{4}}(\beta+q^{2}u_{0}\cos qy),\label{eq:III_velocity}\\
\frac{dp_{y}}{dt}=p_{x}qu_{0}\left(1-\frac{q^{2}}{p_{D}^{2}}\right)\sin qy.\label{eq:III_acceleration}
\end{gather}
Due to the assumed GO approximation, we limit our consideration to
the regime where $q^{2}\ll1$ (in dimensional form, $q^{2}\ll\rho_{s}^{-2}$).
We also assume that $p_{x}$ is nonzero. Then, by studying the drifton
phase-space trajectories governed by Eqs.~(\ref{eq:III_velocity})
and (\ref{eq:III_acceleration}), one can identify three distinct
regimes depending on how the ZF magnitude $u_{0}$ compares with the
two critical values (the derivations are given in Appendix \ref{appendixA}),
\begin{eqnarray}
u_{c,1}\doteq\frac{\beta}{2-q^{2}}, & \quad & u_{c,2}\doteq\frac{\beta}{q^{2}}.\label{eq:III_uc}
\end{eqnarray}
The GO approximation implies $0<u_{c,1}\ll u_{c,2}$. The phase-space
structures are illustrated in Fig.~\ref{fig:III_traj} that shows
typical contour plots of $\mathcal{H}$ corresponding to three distinct
regimes. (In a stationary ZF considered here, driftons travel along
constant-energy surfaces.) Specifically, these three regimes are as
follows.

$\textit{Regime 1.--}$ The first regime corresponds to $u_{0}<u_{c,1}$
(weak ZF). In this regime, there are two types of phase-space stationary
points, namely, the stable stationary points (centers) at
\begin{equation}
y=\pm\frac{\pi}{q},\quad p_{y}=0,\label{eq:III_center}
\end{equation}
and the unstable stationary point (saddle) at
\begin{equation}
y=0,\quad p_{y}=0.\label{eq:III_saddle}
\end{equation}
(Due to the periodicity of the system, we limit our consideration
to a single period, $y\in[-\pi/q,\pi/q]$.) The trajectories in this
regimes are of three different types {[}Fig. \ref{fig:III_traj}(a){]}:
passing (labeled by ``P''), trapped (labeled by ``T''), and runaway
(labeled by ``R''). Passing trajectories reside near the saddle,
while trapped trajectories reside near the centers. Passing and trapped
trajectories are qualitatively similar to those predicted by the tWKE
\cite{Kaw2001,Singh2014,Sasaki2018} and are also reminiscent of the
corresponding trajectories of charged particles interacting with plasma
waves. If these were the only trajectories, a nonlinear ZF\textendash DW
system in this regime would have been similar to a BGK wave \cite{BGK1957},
but the runaway trajectories make the picture qualitatively different.
These trajectories are localized spatially around $y=0$ but extend
to infinity along the momentum axis; therefore, these driftons tend
to $\textit{accumulate}$ in certain spatial locations. (This is understood
from the fact that, while $|p_{y}|$ remains growing, the DW group
velocity at large $|p_{y}|$ is $\dot{y}\propto p_{y}^{-3}\to0$,
so a drifton eventually stops moving along $y$.) The existence of
runaway driftons indicates that contrary to Refs.~\cite{Kaw2001,Singh2014,Sasaki2018},
a ZF--DW system cannot be in an exact steady state; otherwise $F$
is a function of $\mathcal{H}$ only {[}Eq.~(\ref{eq:II_WKE_Parker}){]}
and the runaway trajectories, if populated, will make $W$ non-integrable.

Note that runaway trajectories are possible even for arbitrarily small
$u_{0}$. Also note that the runaway trajectories can also be obtained
from the tWKE {[}Eqs.~(\ref{eq:II_tWKE_hamiltonian}) and (\ref{eq:II_tWKE_dissipation}){]}.
This explains that some plots in Refs.~\cite{Singh2014,Sasaki2018}
obtained from the tWKE look similar to our Fig.~\ref{fig:III_traj}(a),  even though the underlying models are different.

$\textit{Regime 2.--}$ The fraction of passing trajectories shrinks
with the increase of $u_{0}$. In the second regime, when $u_{c,1}\leq u_{0}\leq u_{c,2}$
(moderate ZF), passing trajectories disappear entirely. This is illustrated
in Fig. \ref{fig:III_traj}(b). The centers and the saddle are also
given by Eqs.~(\ref{eq:III_center}) and (\ref{eq:III_saddle}).
The trapped trajectories reside near the center, while the remaining
phase space corresponds to runaways.

Note that Regime~2 is also possible in the tWKE, where the derivatives
of $U$ are omitted in $\mathcal{H}$, so $q^{2}$ is effectively
set to zero. This leads to $u_{c,1}=\beta/2$, which remains finite.
Therefore, passing trajectories also disappear entirely if $u_{0}\geq u_{c,1}$.
On the other hand, setting $q^{2}$ to zero leads to $u_{c,2}=+\infty$,
hence the Regime~3 below is impossible in the tWKE.

$\textit{Regime 3.--}$ The third regime corresponds to $u_{0}\geq u_{c,2}$
(strong ZF). In this regime, all the stationary points {[}Eqs.~(\ref{eq:III_center})
and (\ref{eq:III_saddle}){]} are unstable, so trapped trajectories
also disappear, and only runaway trajectories are left. As is illustrated
in Fig.~\ref{fig:III_traj}(c), all the driftons move towards
\begin{gather}
|y|=y_{*},\quad|p_{y}|=\infty.
\end{gather}
Here, $\pm y_{*}$ are the locations where $U''=\beta$; namely,
\begin{equation}
y_{*}=\frac{1}{q}\left[\pi-\arccos\left(\frac{\beta}{q^{2}u_{0}}\right)\right].\label{eq:III_ystar}
\end{equation}
In particular, note that no trajectory can cross the vertical lines
$|y|=y_{*}$ {[}shown as vertical white dashed lines in Fig.~\ref{fig:III_traj}(c){]},
since the drifton velocity is always zero at $|y|=y_{*}$ {[}Eq.~(\ref{eq:III_velocity}){]}. 

Remarkably, the condition under which Regime~3 is realized (i.e.,
that $u_{0}>u_{c,2}\doteq\beta/q^{2}$) is precisely the Rayleigh\textendash Kuo
criterion \cite{Kuo1949}, which states that a necessary condition
for the ZF instability is the existence of spatial locations where
$U''=\beta$. The connection between Regime~3 and the R\textendash K
criterion is only captured by the improved Hamiltonian {[}Eq.~(\ref{eq:II_WKE_hamiltonian}){]}.
In the tWKE, where the $U''$ term in $\mathcal{H}$ is neglected,
the R\textendash K criterion is never reached, and hence Regime~3
cannot be realized. However, as will be argued below (Sec.~\ref{sec:5}),
Regime~3 in the iWKE does not quite play the same role as that in
the R\textendash K criterion. In contrast with the full-wave theory,
the GO model predicts that driftons in Regime~3 \emph{amplify} a
ZF rather than destroy it, as we will now discuss.

\section{Nonlinear saturation of the zonostrophic instability}

\label{sec:section4}

Here, we study the nonlinear structures of DW turbulence in ZFs based
on the drifton phase-space trajectories presented in Sec.~\ref{sec:section3}.
Specifically, we consider the ZI, which describes the formation of
ZFs out of DW turbulence with a given equilibrium drifton Wigner function
$\mathcal{W}(\boldsymbol{p})$ \cite{Smolyakov2000,Smolyakov2000prl,Srinivasan2012,Parker2013,Parker2014}.
Assuming perturbations of the form $U=\text{Re}\,(U_{q}e^{iqy+\gamma_{\text{ZI}}t})$
and $\delta W=\text{Re}\,(W_{q}e^{iqy+\gamma_{\text{ZI}}t})$, the
linearized Eqs\@.~(\ref{eq:II_WKE_hamiltonian}) and (\ref{eq:II_WKE_dissipation})
are
\begin{multline}
W_{q}=\frac{1}{\gamma_{{\rm ZI}}+2i\beta qp_{x}p_{y}/p_{D}^{4}}\\
\times\left[iqp_{x}\left(1-\frac{q^{2}}{p_{D}^{2}}\right)\frac{\partial\mathcal{W}}{\partial p_{y}}+\frac{2iq^{3}p_{x}p_{y}}{p_{D}^{4}}\,\mathcal{W}\right],\label{eq:IV_linearize_W}
\end{multline}
\begin{gather}
U_{q}=\frac{iq}{\gamma_{{\rm ZI}}}\int\frac{d^{2}p}{(2\pi)^{2}}\,\frac{p_{x}p_{y}}{p_{D}^{4}}\,W_{q}.\label{eq:IV_linearize_U}
\end{gather}
By plugging (\ref{eq:IV_linearize_W}) into (\ref{eq:IV_linearize_U})
and integrating by parts the term that contains $\partial\mathcal{W}/\partial p_{y}$,
we obtain the dispersion relation for the linear ZI of the iWKE \cite{Parker2016,Zhu2017}:
\begin{equation}
1=\int\frac{d^{2}p}{(2\pi)^{2}}\,\frac{q^{2}p_{x}^{2}p_{D}^{4}(1-4p_{y}^{2}/p_{D}^{2})(1-q^{2}/p_{D}^{2})}{\left(\gamma_{\text{ZI}}p_{D}^{4}+2i\beta qp_{x}p_{y}\right)^{2}}\,\mathcal{W}(\boldsymbol{p}).\label{eq:IV_gamma_ZI}
\end{equation}

\begin{figure}
\includegraphics[width=1\columnwidth]{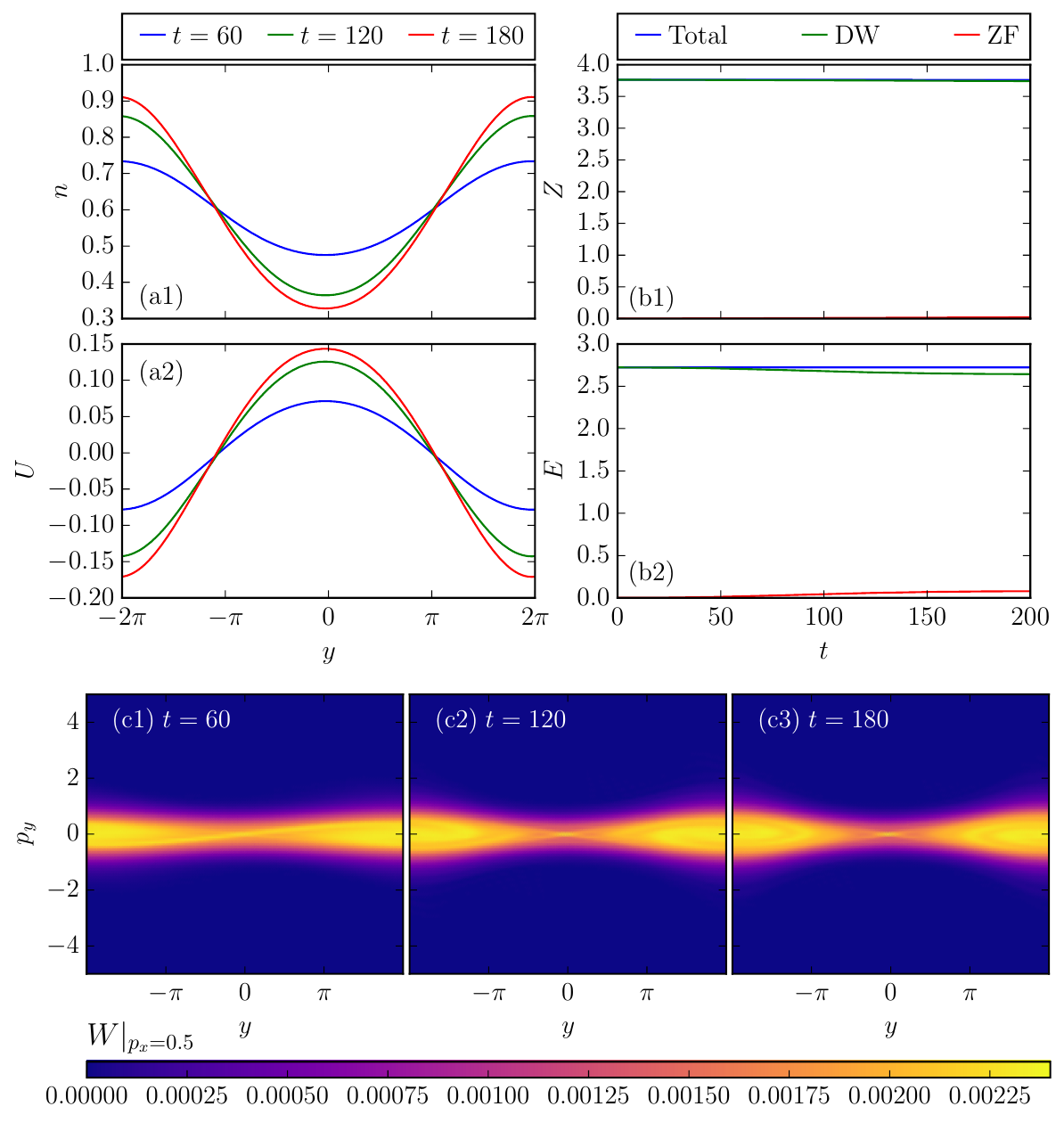}
\caption{Results of iWKE simulations with initial conditions given by Eqs.~(\ref{eq:IV_mc(W)})
and (\ref{eq:IV_U(0)}). Here, the parameters are $\beta=1$, $q=0.5$,
$W_{0}=0.3$, $r=0.5$, $U_{q}=0.01$, and the hyperviscocity coefficient
is $\nu=1\times10^{-7}$ \cite{Parker2014}. (a) The drifton density
$n$ {[}Eq.~(\ref{eq:IV_fluid_n}){]} and the ZF velocity $U$ at
$t=60$, $120$, and $180$. The relation between the change of $n$
and $U$ agrees with Eq.~(\ref{eq:IV_fluid_final}). (b) The time
evolution of the energy $E$ and enstrophy $Z$ integrated over one spatial
period $y\in[-2\pi,2\pi]$. The energy and enstrophy exchange between
DWs and ZF is small, since the ZF is weak. (c) The drifton phase-space
Wigner function $W(y,p_{y})$ at $p_{x}=0.5$ at the three different
instants. Passing and trapped trajectories are clearly seen. [Associated dataset available at: \url{http://dx.doi.org/10.5281/zenodo.1244318}] \cite{dataset} \label{fig:III_regime1}}
\end{figure}

\begin{figure}
\includegraphics[width=1\columnwidth]{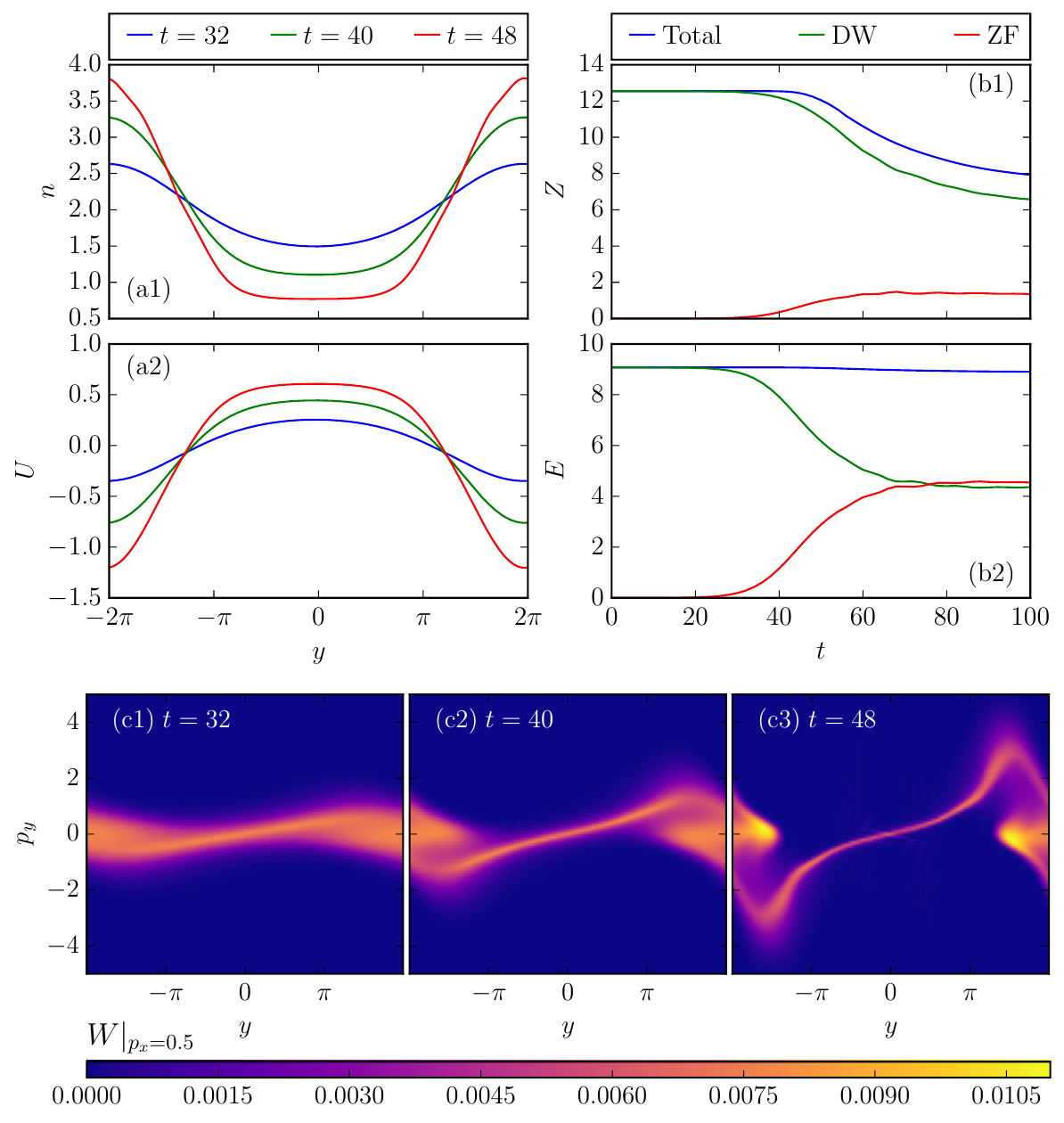}
\caption{The same as Fig.~\ref{fig:III_regime1}, except with $W_{0}=1$ and $\nu=2\times10^{-7}$.
In (b), the decrease of the total enstrophy at $t\gtrsim40$ is due
to the fact that runaway driftons at large $|p_{y}|$ are heavily
damped by hyperviscosity. It is seen that the energy exchange between
DWs and ZF is large due to runaways. In (c), trapped and runaway trajectories
are clearly seen. [Associated dataset available at: \url{http://dx.doi.org/10.5281/zenodo.1244318}] \cite{dataset} \label{fig:III_regime2}}

\end{figure}

\begin{figure}
\includegraphics[width=1\columnwidth]{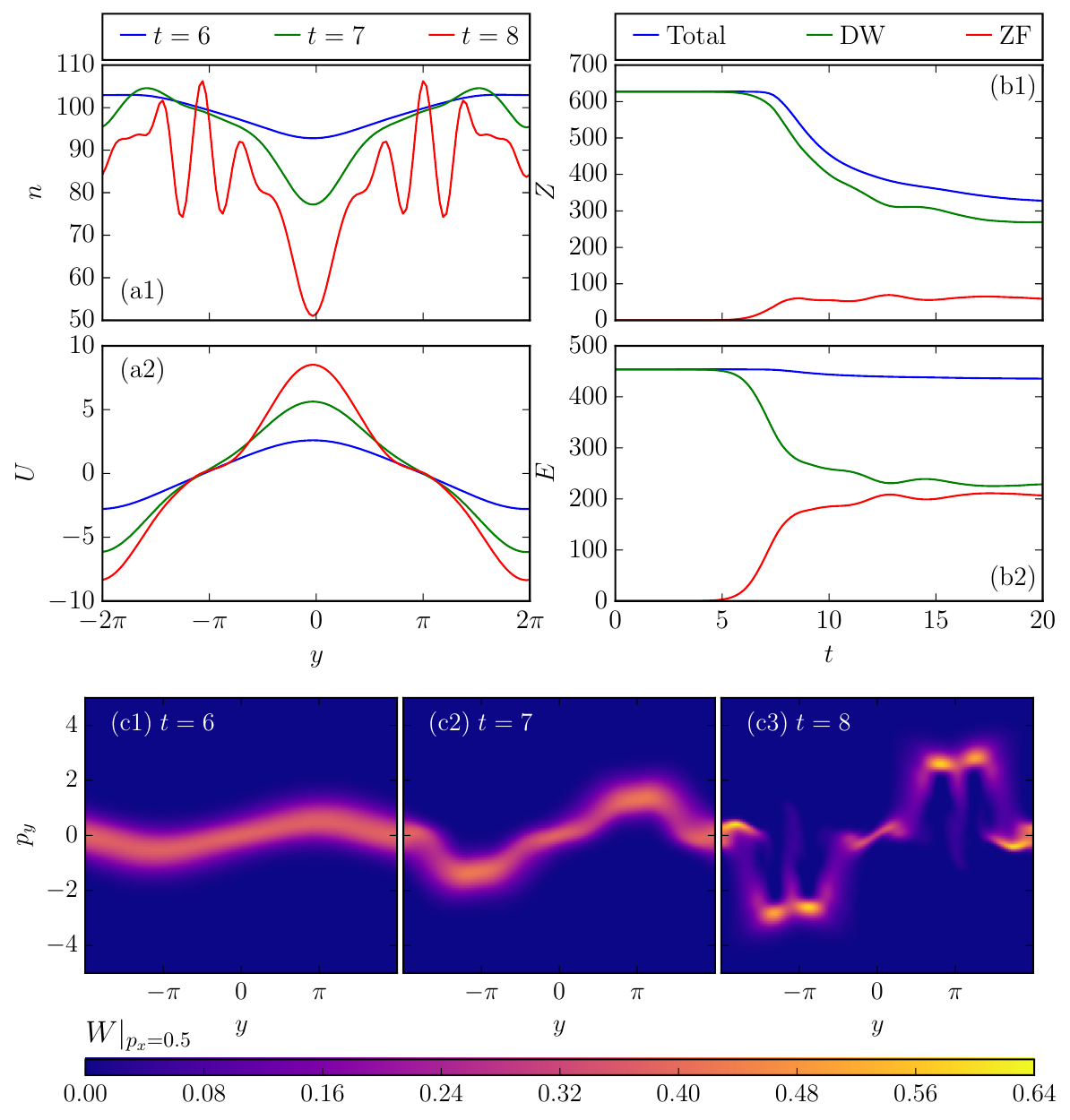}
\caption{The same as Figs.~\ref{fig:III_regime1} and \ref{fig:III_regime2},
except with $W_{0}=50$ and $\nu=10^{-5}$. In (b), the decrease of the
total enstrophy at $t\gtrsim7$ is due to the fact that runaway driftons
at large $|p_{y}|$ are heavily damped by hyperviscosity. It is seen
that the energy exchange between DWs and ZF is large due to the runaways.
Note that the ZF fails to absorb the whole DW energy, which is explained
in the text. In (c), runaway trajectories are clearly seen. However,
more intricate structures emerge too, because the ZF is far from sinusoidal. [Associated dataset available at: \url{http://dx.doi.org/10.5281/zenodo.1244318}] \cite{dataset}
\label{fig:III_regime3}}

\end{figure}

As the ZF amplitude becomes finite, the ZI enters its nonlinear stage
and eventually saturates. Previous studies discussed how the saturated
state is determined by the interplay of ZFs and passing and trapped
orbits \cite{Kaw2001,Singh2014,Sasaki2018,Li2018}. However, this
picture is qualitatively altered by runaway trajectories. To show
this, we numerically simulate the iWKE {[}Eqs\@.~(\ref{eq:II_WKE_hamiltonian})
and (\ref{eq:II_WKE_dissipation}){]} using the pseudo-spectral method
described in Ref.~\cite{Parker2018} (where the iWKE is termed ``CE2-GO'').
A weak 8th-order hyperviscosity is added for numerical stability \cite{Parker2014}.
We launch the simulation with an initial Gaussian DW distribution
\begin{equation}
\mathcal{W}(\boldsymbol{p})\doteq W(t=0,y,\boldsymbol{p})=\frac{4\pi W_{0}}{r^{2}}\exp\left(-\frac{|\boldsymbol{p}|^{2}}{2r^{2}}\right)\label{eq:IV_mc(W)}
\end{equation}
(where $r$ is some constant serving as a characteristic DW wavenumber)
and an initial ZF perturbation
\begin{equation}
U(t=0)=U_{q}\cos qy\label{eq:IV_U(0)}
\end{equation}
with small $U_{q}$. It is found that depending on the strength of
the DW amplitude $W_{0}$ {[}Eq.~(\ref{eq:IV_mc(W)}){]}, the ZF
can saturate in one of the three different regimes described in Sec.~\ref{sec:section3}.
The simulation results are shown in Figs.~\ref{fig:III_regime1},
\ref{fig:III_regime2}, and \ref{fig:III_regime3}, where the structure
of the DW Wigner functions $W$ reflects the structure of the underlying
drifton trajectories.

At the nonlinear stage of ZI, the time-evolution of ZFs can be qualitatively
estimated from the drifton trajectories. In order to demonstrate this,
we derive the drifton hydrodynamic equations as follows. Let us write
down Eq.~(\ref{eq:II_WKE_DW}) explicitly:
\begin{multline}
\frac{\partial W}{\partial t}=\left(U'''\frac{p_{x}}{p_{D}^{2}}+U'p_{x}\right)\frac{\partial W}{\partial p_{y}}\\
+\frac{2(U''-\beta)p_{x}p_{y}}{p_{D}^{4}}\,\frac{\partial W}{\partial y}-\frac{2p_{x}p_{y}U'''}{p_{D}^{4}}\,W.\label{eq:IV_WKE}
\end{multline}
Integrating Eq.~(\ref{eq:IV_WKE}) over $\boldsymbol{p}$ leads to
the following equation for the drifton density $n$:
\begin{equation}
\frac{\partial n}{\partial t}+(\beta-U'')\,\frac{\partial V}{\partial y}=0,\label{eq:IV_fluid_n}
\end{equation}
where
\begin{eqnarray}
n(y,t) & \doteq & \int\frac{d^{2}p}{(2\pi)^{2}}\,W(y,\boldsymbol{p},t),\label{eq:IV_fluid_definition_n}\\
V(y,t) & \doteq & \int\frac{d^{2}p}{(2\pi)^{2}}\,\frac{2p_{x}p_{y}}{p_{D}^{4}}\,W(y,\boldsymbol{p},t).\label{eq:IV_fluid_definition_V}
\end{eqnarray}
Then, one can rewrite Eq.~(\ref{eq:II_WKE_ZF}) as follows:
\begin{equation}
\frac{\partial U}{\partial t}=\frac{1}{2}\,\frac{\partial V}{\partial y},\label{eq:IV_fluid_U}
\end{equation}
and the combination of Eqs.~(\ref{eq:IV_fluid_definition_n}) and
(\ref{eq:IV_fluid_definition_V}) gives
\begin{equation}
\frac{\partial U}{\partial t}=\frac{1}{2(U''-\beta)}\,\frac{\partial n}{\partial t}.\label{eq:IV_fluid_final}
\end{equation}
[Even though the denominator is zero at $U''=\beta$, the right-hand
side of Eq.~(\ref{eq:IV_fluid_final}) remains finite because, according
to Eqs.~(\ref{eq:IV_fluid_n}), $\partial n/\partial t$ is also
zero at such locations.] 

By using the knowledge of the phase-space trajectories, which determine
the drifton flows, one can predict whether the drifton density grows
or decreases at a given location. This gives the sign of $\partial n/\partial t$;
then, the sign of $\partial U/\partial t$ can be inferred from Eq.
(\ref{eq:IV_fluid_final}), so one can tell whether the ZF is peaking
or flattening. In Regimes~1 and 2, $U''-\beta$ is always negative;
hence, $\partial U/\partial t$ and $\partial n/\partial t$ have opposite
signs, which is consistent with Figs.~\ref{fig:III_regime1} and
\ref{fig:III_regime2}. Regime~3 is more interesting due to its connection
with the R\textendash K criterion. Let us consider a ZF of the assumed
sinusoidal form [Eq.~(\ref{eq:III_U_profile})] that satisfies
$u_{0}>u_{c,2}$. Then, from Fig.~\ref{fig:III_traj}(c), it is seen
that within $|y|<y_{*}$, driftons move away from $y=0$. Therefore,
$n$ decreases at $y=0$. Since $U''-\beta<0$ at $y=0$, from Eq.~(\ref{eq:IV_fluid_final})
we have
\begin{equation}
\frac{\partial U(y=0)}{\partial t}>0.
\end{equation}
A similar argument leads to
\begin{equation}
\frac{\partial U(y=\pm\pi/q)}{\partial t}<0.
\end{equation}
Hence, the ZF profile gets more peaked, i.e., the ZF is $\textit{globally amplified}$.
The amplification of the ZF, in turn, will reinforce the drifton runaway.
From the expression of the DW energy {[}Eq.~(\ref{eq:II_energy_dw}){]},
the bulk motion of driftons to $|p_{y}|=\infty$ causes $p_{D}^{2}\to\infty$;
hence $E_{{\rm DW}}\to0$. Therefore, in Regime~3, the ZF tends to
absorb \textit{all} the energy from DWs; The reasons why this does
not happen in Fig.~\ref{fig:III_regime3} are twofold: (i) in
order to better visualize drifton trajectories, we have chosen a Gaussian
initial distribution centered at $\boldsymbol{p}=0$ {[}Eq.~(\ref{eq:IV_mc(W)}){]};
hence a large fraction of driftons with $p_{x}\ll1$, which move in
phase space slowly {[}Eqs.~(\ref{eq:II_ray_velocity}) and (\ref{eq:II_ray_acceleration}){]},
does not significantly participate in the energy exchange; also, (ii)
some driftons do not runaway as can be seen in Fig.~\ref{fig:III_regime3}(c),
since the ZF is far from sinusoidal in this highly nonlinear stage.

\section{There is no tertiary instability in the GO limit}

\label{sec:section5}

In addition to the ZI of a ZF\textendash DW system, which leads to
the ZF amplification, it is also of interest to examine whether the
iWKE is applicable to describe the so-called tertiary instability
(TI) \cite{Kim2002,Rath2018,Rogers2000,Rogers2005,Numata2007,St-Onge2017,Singh16}. Specifically, we define the TI as the instability of a DW on top of a prescribed non-turbulent ZF equilibrium, i.e., an instability of
a Kelvin--Helmholtz type. Note that this definition is different from
that in Refs.~\cite{Rogers2000,Rogers2005}, where the TI was attributed
to the ion-temperature gradient (absent in our model), but similar
to that in the majority of relevant papers \cite{Kim2002,Numata2007,St-Onge2017,Singh16}.

As speculated in Refs.~\cite{Parker2016,Numata2007} and later elaborated
in Ref.~\cite{Zhu2017}, the Rayleigh\textendash Kuo criterion is a necessary
condition for this instability, so one might expect the TI to develop
in Regime~3. However, as shown above, ZF can only grow in Regime~3
rather than deteriorate, so in principle, there is no TI in the WKE
under the GO assumption. The reason is that, as shown in Ref.~\cite{Zhu2018},
within the gHME, the TI requires $q^{2}>1$, while a GO model relies
on the assumption that $q^{2}\ll1$. 

Another explanation for the absence of the TI in our model is as follows.
Let us consider a small DW perturbation around a stationary ZF. Linearizing
Eq.~(\ref{eq:II_WKE_DW}) gives
\begin{equation}
\frac{\partial W_{1}}{\partial t}=\{\mathcal{H}_{0},W_{1}\}+2\Gamma_{0}W_{1},\label{eq:V_linearization}
\end{equation}
where $W_{1}$ is the Wigner function of driftons. (The zeroth-order
DW Wigner function is zero because, as mentioned earlier, we assume
a non-turbulent background for the TI.) Since $\mathcal{H}_{0}$ and
$\Gamma_{0}$ are stationary, Eq.~(\ref{eq:V_linearization}) is
equivalent to
\begin{equation}
\frac{d}{dt}\left(\frac{W_{1}}{U''-\beta}\right)=0
\end{equation}
(as we mentioned in Sec.~\ref{subsec:section2.3} for a similar equation),
where $d/dt$ is taken along the drifton trajectories determined by
Eqs.~(\ref{eq:II_ray_velocity}) and (\ref{eq:II_ray_acceleration}).
As shown in Appendix \ref{appendixA}, drifton runaway trajectories do not reach locations
where $U''=\beta$. Hence, $U''-\beta$ remains finite along trajectories,
and $W_{1}$ cannot grow exponentially with time. This rules out the
TI and, accordingly, this also means that the TI cannot be described
by the quasilinear WKE, because the WKE relies on the GO approximation. (However, full-wave quasilinear models are perfectly capable of capturing the TI; for example, see Refs.~\cite{Zhu2017,Zhu2018,Marston2016}.)

\section{Conclusions}

\label{sec:5}

In summary, this paper presents the first study of the drifton phase-space
dynamics within the iWKE proposed in Refs.~\cite{Parker2016,Ruiz2016}.
Contrary to the traditional GO model of the drifton dynamics, it is
found that driftons can be not only trapped or passing, but they can
also accumulate spatially while experiencing indefinite growth of
their momenta. In particular, it is found that the Rayleigh\textendash Kuo
threshold known from geophysics corresponds to the regime when such
``runaway'' trajectories are the only ones possible. On one hand,
this analysis helps visualize the development of the ZI, particularly
its nonlinear stage, which we study both analytically and through
iWKE simulations. On the other hand, the GO theory predicts that ZFs
above the Rayleigh\textendash Kuo threshold can only grow; hence,
the deterioration of intense ZFs cannot be captured within a GO model.
In particular, this means that the so-called tertiary instability
of intense zonal flows cannot be adequately described within the quasilinear
WKE, contrary to some previous studies. 
\begin{acknowledgments}
The authors thank J.~B.~Parker for providing a copy of his code
for our wave-kinetic simulations. This work was supported by the U.S.
Department of Energy (DOE), Office of Science, Office of Basic Energy
Sciences, and also by the U.S. DOE through Contract No. DE-AC02-09CH11466.
\end{acknowledgments}

\appendix

\section{Derivation of drifton trajectories}

\label{sec:appendix A}
\label{appendixA}

Here, we give a detailed description of the drifton phase-space trajectories
governed by Eqs.~(\ref{eq:III_velocity}) and (\ref{eq:III_acceleration}).
We also derive the two critical ZF magnitudes given by Eq.~(\ref{eq:III_uc}).
Assuming a sinusoidal ZF {[}Eq.~(\ref{eq:III_U_profile}){]}, the
drifton Hamiltonian can be written as
\begin{multline}
\mathcal{H}(y,p_{y})=p_{x}u_{0}\cos qy-\frac{p_{x}}{1+p_{x}^{2}+p_{y}^{2}}\left(\beta+q^{2}u_{0}\cos qy\right),\label{eq:A_H}
\end{multline}
where $p_{x}$ is a constant parameter, and we assume $p_{x}>0$ without
loss of generality. Since driftons move along constant-$\mathcal{H}$
surfaces, the drifton trajectory can be determined by equating $\mathcal{H}(y,p_{y})$
to a constant $\mathcal{E}$, where $\mathcal{E}\doteq\mathcal{H}(y_{0},p_{y_{0}})$
is determined by the initial location in the phase space. Then, we
obtain $p_{y}$ as a function of $y$: 
\begin{eqnarray}
p_{y}^{2}(y) & = & (1+p_{x}^{2})\,\frac{\mathcal{E}-\mathcal{H}^{0}(y)}{\mathcal{H}^{\infty}(y)-\mathcal{E}},\label{eq:A_py_y}
\end{eqnarray}
where we introduced
\begin{multline}
\mathcal{H}^{0}(y)\doteq\mathcal{H}(y,p_{y}=0)\\
=p_{x}u_{0}\cos qy-\frac{p_{x}}{1+p_{x}^{2}}\left(\beta+q^{2}u_{0}\cos qy\right),\label{eq:A_H0}
\end{multline}
and
\begin{gather}
\mathcal{H}^{\infty}(y)\doteq\mathcal{H}(y,p_{y}=\infty)=p_{x}u_{0}\cos qy.\label{eq:A_Hinf}
\end{gather}
It is straightforward to show that $\max\mathcal{H}^{0}=\mathcal{H}^{0}(y=0)$
and $\min\mathcal{H}^{0}=\mathcal{H}^{0}(y=\pm\pi/q)$ (and the same
for $\mathcal{H}^{\infty}$); one can also quickly show that the range
of $\mathcal{E}$ is given by 
\begin{equation}
\mathcal{E}\doteq\mathcal{H}(y_{0},p_{y_{0}})\in\left[\min(\min\mathcal{H}^{\infty},\min\mathcal{H}^{0}),\max\mathcal{H}^{\infty}\right],\label{eq:A_E_range}
\end{equation}
where $\mathcal{E}=\max\mathcal{H}^{\infty}$ is achieved at $(y_{0}=0,p_{y_{0}}=\infty)$,
$\mathcal{E}=\min\mathcal{H}^{\infty}$ is achieved at $(y_{0}=\pm\pi/q,p_{y_{0}}=\infty)$,
and $\mathcal{E}=\min\mathcal{H}^{0}$ is achieved at $(y_{0}=\pm\pi/q,p_{y_{0}}=0)$.

\begin{figure}[H]
\includegraphics[width=1\columnwidth]{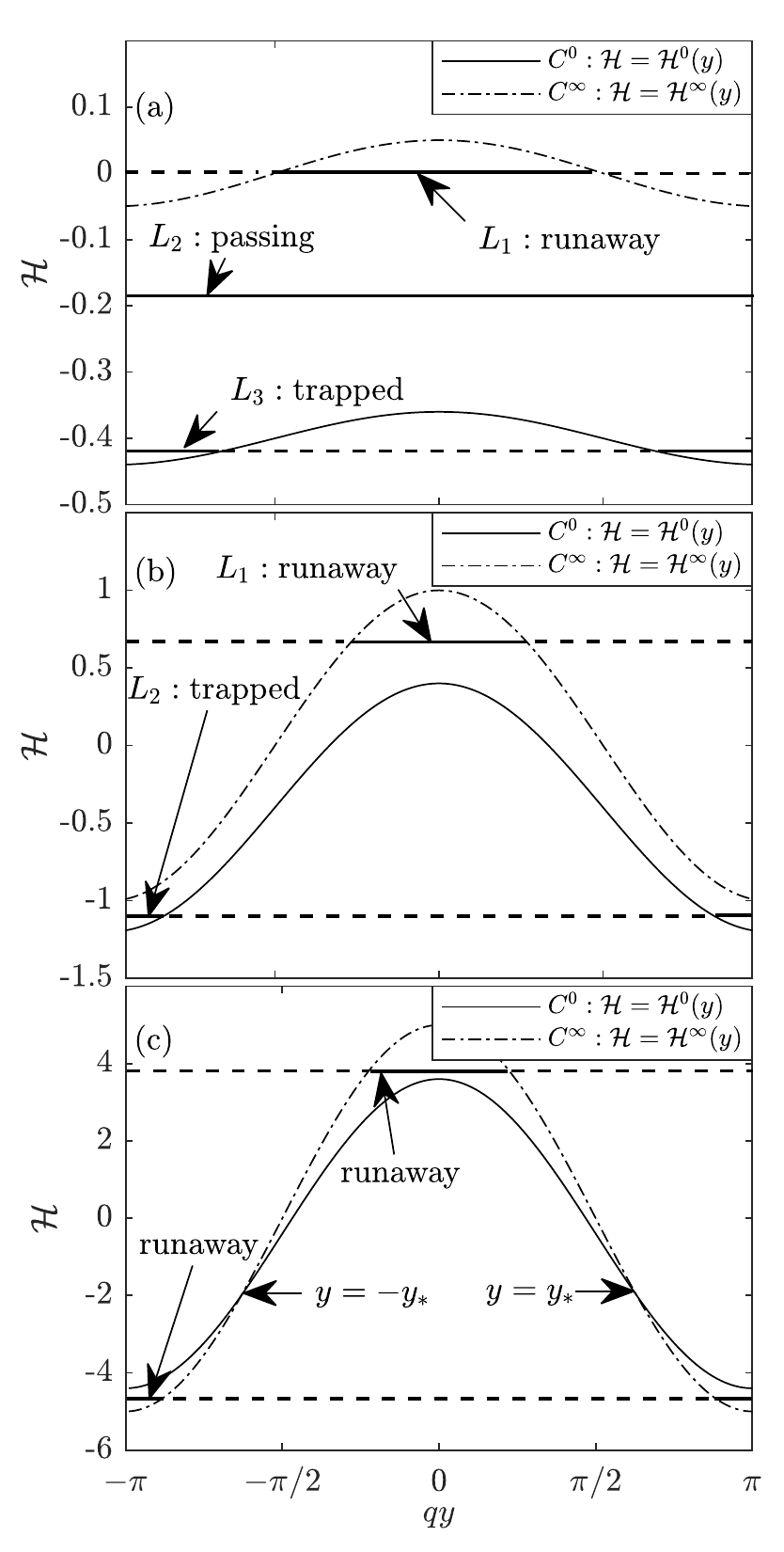}
\caption{$\mathcal{H}=\mathcal{H}^{0}(y)$ (solid curves), $\mathcal{H}=\mathcal{H}^{\infty}(y)$
(dot-dashed curves), and $\mathcal{H}=\mathcal{E}$ (horizontal lines)
for (a) $u_{0}=0.1$ (Regime~1), (b) $u_{0}=2$ (Regime~2), and
(c) $u_{0}=10$ (Regime~3). In all figures, the parameters are $\beta=1$,
$q=0.5$, $p_{x}=0.5$. The definitions of $\mathcal{H}^{0}(y)$ and
$\mathcal{H}^{\infty}(y)$ are given by Eqs.~ (\ref{eq:A_H0}) and
(\ref{eq:A_Hinf}). Each horizontal line represents a drifton trajectory.
Note that drifton trajectories are confined within the spatial regions
indicated by the solid-line parts of these horizontal lines. \label{fig:A_4_trajectory}}
\end{figure}

The drifton trajectories can be conveniently studied on the $(y,\mathcal{H})$
plane using the following method. We plot two curves $C^{0}:\mathcal{H}=\mathcal{H}^{0}(y)$
and $C^{\infty}:\mathcal{H}=\mathcal{H}^{\infty}(y)$ (Fig.~\ref{fig:A_4_trajectory})
and draw a horizontal line $L$ that represents $\mathcal{H}=\mathcal{E}$
{[}note that $\mathcal{E}$ should be within the range (\ref{eq:A_E_range}){]}.
If $L$ intersects neither $C^{0}$ nor $C^{\infty}$, then $p_{y}(y)$
is always finite, hence the trajectory is passing. If $L$ intersects
$C^{0}$ at some location, then $p_{y}(y)=0$. But $dp_{y}/dt$ is
nonzero according to Eq.~(\ref{eq:III_acceleration}); hence, the
drifton will bounce back at that location, which indicates a trapped
trajectory (provided that $L$ does not intersect $C^{\infty}$).
If $L$ intersects $C^{\infty}$ at some location, then $|p_{y}(y)|=\infty$,
which indicates that the drifton is running away in $p_{y}$ space
while approaching a particular spatial location. (However, the drifton
will never reach such locations, because $dp_{y}/dt$ is finite everywhere.)
This indicates a runaway trajectory. 

For illustration purpose, let us plot $C^{0}$ and $C^{\infty}$ in
Fig.~\ref{fig:A_4_trajectory} for different values of $u_{0}$ and
study the corresponding trajectories. Figure \ref{fig:A_4_trajectory}(a)
is for $u_{0}=0.1$, which corresponds to Regime~1; $L_{1}$, $L_{2}$,
and $L_{3}$ drawn there correspond to runaway, trapped, and passing
trajectories. Figure \ref{fig:A_4_trajectory}(b) is for $u_{0}=2$,
which corresponds to Regime~2; $L_{1}$ and $L_{2}$ drawn there
correspond to runaway and trapped trajectories, while no $L$ can
simultaneously avoid intersecting both $C^{0}$ and $C^{\infty}$,
hence no passing trajectory exists. Figure \ref{fig:A_4_trajectory}(c)
is for $u_{0}=10$, which corresponds to Regime~3; in this case,
$C^{0}$ and $C^{\infty}$ intersect at $|y|=y_{*}$ {[}Eq.~(\ref{eq:III_ystar}){]},
hence every $L$ intersects $C^{\infty}$, giving a runaway trajectory.
Note that the horizontal lines in Fig.~\ref{fig:A_4_trajectory}
are divided into solid-line parts and dashed-line parts; the actual
drifton trajectories are confined in the spatial regions indicated
by the solid-line parts.

More formally, this method of classifying trajectories can also be
presented as follows. For a passing trajectory, $L$ intersects neither
$C^{0}$ nor $C^{\infty}$; in other words, if the criterion
\begin{equation}
P:\mathcal{E}>\max\mathcal{H}^{0}\,\,{\rm and}\,\,\mathcal{E}<\min\mathcal{H}^{\infty}\label{eq:A_P}
\end{equation}
is satisfied, then we have a passing trajectory. For a trapped trajectory,
$L$ intersects $C^{0}$ but not $C^{\infty}$, hence the corresponding
criterion is
\begin{equation}
T:\min\mathcal{H}^{0}\leq\mathcal{E}\leq\max\mathcal{H}^{0}\,\,{\rm and}\,\,\mathcal{E}<\min\mathcal{H}^{\infty}.\label{eq:A_T}
\end{equation}
For a runaway trajectory, $L$ intersects $C^{\infty}$, hence the
corresponding criterion is
\begin{equation}
R:\min\mathcal{H}^{\infty}\leq\mathcal{E}\leq\max\mathcal{H}^{\infty}.\label{eq:A_R}
\end{equation}
Consequently, the conditions for each type of trajectory to exist
are given as follows. First, passing trajectories exist when $\max\mathcal{H}^{0}<\min\mathcal{H}^{\infty}$
(otherwise, the criterion $P$ is never satisfied for any $\mathcal{E}$);
this gives
\begin{equation}
u_{0}<\frac{\beta}{2(1+p_{x}^{2})-q^{2}}\leq u_{c,1}\doteq\frac{\beta}{2-q^{2}},\label{eq:A_uc1}
\end{equation}
where the equality sign applies at $p_{x}=0$. Next, trapped trajectories
exist when $\min\mathcal{H}^{0}<\min\mathcal{H}^{\infty}$ (otherwise,
the criterion $T$ is never satisfied for any $\mathcal{E}$); this
gives
\begin{equation}
u_{0}<u_{c,2}\doteq\frac{\beta}{q^{2}}.\label{eq:A_uc2}
\end{equation}
Note that under the GO assumption $q^{2}\ll1$, we have $u_{c,1}\ll u_{c,2}$.
Finally, runaway trajectories always exist, since one can always find
$\mathcal{E}$ within the range (\ref{eq:A_E_range}) that satisfy
the criterion $R$.

The above criteria help us quickly identify the three regimes as follows.
If $u_{0}<u_{c,1}$, trapped and runaway trajectories exist; passing
trajectories also exist because $u_{0}$ satisfies Eq.~(\ref{eq:A_uc1})
for small enough $p_{x}$, hence we have Regime~1. If $u_{c,1}\leq u_{0}<u_{c,2}$,
only trapped and runaway trajectories exist; hence we have Regime~2.
If $u_{0}\geq u_{c,2}$, only runaway trajectories exist, hence we
have Regime~3.

We also emphasize that runaway trajectories are possible even in the
tWKE {[}Eq.~(\ref{eq:II_tWKE_hamiltonian}){]}, which is equivalent
to setting $q^{2}$ to zero. In this case, the two critical ZF magnitudes
become $u_{c,1}=\beta/2$ and $u_{c,2}=+\infty$. Then, $u_{0}<u_{c,2}$
is satisfied automatically, and the system is always either in Regime~1
or in Regime~2. However, the criterion $R$ can still be satisfied
even after setting $q^{2}$ to zero, so runaway trajectories are still
possible. Moreover, following the same argument as above, we find
that, when $u_{0}\geq u_{c,1}$, passing trajectories also disappear
entirely.

\end{document}